\documentstyle[prd,aps,eqsecnum,preprint,tighten,floats]{revtex}

\def\ha{{1\over 2}} \def\fr#1,#2{{#1\over #2}}
\def\normord#1{\mathopen{\hbox{\bf:}}#1\mathclose{\hbox{\bf:}}}
\def\ran{\rangle} 
\def\ket#1{|#1\ran} 
\def\dots{\ldots} 
 
\def\lan{\langle} 
\def\bra#1{\lan#1|} 
\def\ub#1{\underline{#1}} 
\def\ra{\rightarrow} 
\def\e{\epsilon}

\begin{document}
\draft

\preprint{\vbox{\hfill SMUHEP/96--07 \\
          \vbox{\vskip0.5in}
          }}

\title{Some Lessons from the Schwinger Model}

\author{Gary McCartor}
\address{Department of Physics, Southern Methodist University, Dallas,
TX 75275}

\date{September 17, 1996}

\maketitle

\begin{abstract}
I shall recall a number of solutions to the Schwinger model in
different gauges, having different boundary conditions and using
different quantization surfaces.  I shall discuss various properties
of these solutions emphasizing the degrees of freedom necessary to
represent the solution, the way the operator products are defined and
the effects these features have on the chiral condensate.

\end{abstract}

\pacs{ }

\section{ Introduction }

In this talk I shall discuss the following solutions to the Schwinger
model:

\begin{enumerate}

\item 
The solution of Lowenstein and Swieca$^1$ in Lorentz gauge and in the
continuum (that is, no periodicity conditions imposed).  That solution
has a chiral condensate given by:
$$
\bra{\Omega} \bar{\psi} \psi \ket{\Omega} = \fr{m},{2\pi} e^\gamma cos\theta
$$
\item 
The solution of Nakawaki$^2$ in Coulomb gauge and with antiperiodic
boundary conditions on the surface $t = 0$.  That solution has a
chiral condensate given by:
$$
\bra{\Omega} \bar{\psi} \psi \ket{\Omega} = \fr{1},{2L} e^X cos\theta , 
$$
which in the large $L$ limit goes to:
$$
\bra{\Omega} \bar{\psi} \psi \ket{\Omega} = \fr{m},{2\pi} e^\gamma cos\theta
$$

\item 
The solution I gave$^3$ in light-cone($ \partial_-A^+ = 0$) gauge
with antiperiodic boundary conditions on $x^+ = 0$ (for $\psi_+$) and
antiperiodic boundary conditions on $x^- = 0$ (for $\psi_-$).  That
solution has a chiral condensate given by:
$$
\bra{\Omega} \bar{\psi} \psi \ket{\Omega} = \fr{1},{L} cos\theta ,
$$
which in the large $L$ limit goes to 0.

\item 
The solution of Nakawaki$^4$ in light-cone gauge in the continuum.
That solution has a chiral condensate given by:
$$
\bra{\Omega} \bar{\psi} \psi \ket{\Omega} = \fr{m},{2\pi} e^\gamma cos\theta
$$

\item 
The solution of Vianello$^5$ in light-cone $(A^+=0)$ gauge with
antiperiodic boundary conditions on the surface $t = 0$.  That
solution has a chiral condensate given by:
$$
\bra{\Omega} \bar{\psi} \psi \ket{\Omega} = \fr{1},{2L} e^X cos\theta , 
$$
which in the large $L$ limit goes to:
$$
\bra{\Omega} \bar{\psi} \psi \ket{\Omega} = \fr{m},{2\pi} e^\gamma cos\theta
$$
\end{enumerate}
A principal focus of our discussion will be how operator products are
regulated in these solutions and the effect of that regulation on the
chiral condensate.

\section{ Chiral Condensate }

Let us first review, in some detail, how the chiral condensate arises.
We shall consider Nakawaki's$^2$ Coulomb gauge solution---that is
probably the most straightforward case.

To produce that solution the $\psi$- fields are initialized on $t=0$
as isomorphic to free, massless Fermi fields:
$$
\psi_1(x) = {1\over\sqrt{2L}}e^{-\lambda_1^{(-)}(x)}
\sigma_1(x) e^{-\lambda_1^{(+)}(x)}
$$
$$
\psi_2(x) = {1\over\sqrt{2L}}e^{-\lambda_2^{(-)}(x)}
\sigma_2(x) e^{-\lambda_2^{(+)}(x)}
$$
Where:
$$
\lambda_1^{(+)} = -\sum_{n=1}^\infty {1\over n} 
D(n)e^{-iq(n)x} \quad ; \quad \lambda_1^{(-)} = \sum_{n=1}^\infty 
{1\over n} D^*(n)e^{iq(n)x}
$$
$$
\lambda_2^{(+)} = -\sum_{n=1}^\infty {1\over n} 
C(n)e^{-iq(-n)x} \quad ; \quad \lambda_2^{(-)} = 
\sum_{n=1}^\infty {1\over n} C^*(n)e^{iq(-n)x},
$$
the $C$'s and $D$'s are the fusion operators associated with
Bosonizing the free massless Fermi field, the $\sigma$'s are spurion
operators, and $q(n)={{n\pi}\over {L}}$.  Of interest later will be
the states which are destroyed by the positive frequency fusion
operators.  Defining:
\begin{eqnarray}
\ket{M,N}&=&\delta^*\Bigl(M\Bigr)\dots \delta^*\Bigl(1\Bigr)
	d^*\Bigl(N\Bigr)\dots d^*\Bigl(1\Bigr)\ket0 \qquad(M>0,N>0)
\nonumber \\ \nonumber \ket{M,N}&=&\beta^*\Bigl(M\Bigr)\dots
\beta^*\Bigl(1\Bigr)
	d^*\Bigl(N\Bigr)\dots d^*\Bigl(1\Bigr)\ket0
\qquad(M<0,N>0)\nonumber \\ \ket{M,N}&=&\delta^*\Bigl(M\Bigr)\dots
\delta^*\Bigl(1\Bigr)
	b^*\Bigl(N\Bigr)\dots b^*\Bigl(1\Bigr)\ket0
\qquad(M>0,N<0)\nonumber \\ \nonumber
\ket{M,N}&=&\beta^*\Bigl(M\Bigr)\dots \beta^*\Bigl(1\Bigr)
	b^*\Bigl(N\Bigr)\dots b^*\Bigl(1\Bigr)\ket0 \qquad(M<0,N<0)
\end{eqnarray}
we have:
$$
       (C~{\rm or}~ D)\ket{M,N} = 0
$$
and no other states have this property.  In this gauge, with these
boundary conditions, there is a single degree of freedom in the field
$A^1$ ( independent of $x^1$ ).  It is convenient to define:
$$
            A_1 = {1\over\sqrt{2\pi}} ( \alpha_1 +  \alpha_1^*)
$$
$$
    \partial_0A_1 = { \sqrt{\pi}\over {2iL}} ( \alpha_1 -  \alpha_1^*)
$$
where the $\alpha$'s satisfy the commutation relations for a single
Bose mode.  With these definitions we can calculate the Hamiltonian:
\begin{eqnarray}
H&=&\fr{\pi},{4Le^2}(Q^2 + Q^2_5) + \fr{\pi},{L}
\sum_{n=1}^\infty\left(C^*(n)C(n) + D^*(n)D(n)\right) \nonumber \\
& &+\fr{e^2},{4L} \sum_{n=1}^\infty
\fr{1},{q_1^2(n)}\normord{ 
\bigl[D(n) + C^*(n)\bigr] \bigl[C(n)
+ D^*(n)\bigr] + \bigl[C(n) + D^*(n)\bigr] \bigl[D(n) +
C^*(n)\bigr]
} \nonumber \\ 
& &+\normord{L(\partial_0A^1)^2 -A_1 Q_5 + \fr{e^2},{\pi}LA_1^2}
\nonumber
\end{eqnarray} 
where $Q$ and $Q_5$ are respectively the charge and the pseudocharge.

To diagonalize the Hamiltonian we define a set of operators {$a(n)$}
implicitly through the relations ( $n\neq 0$ ):
$$
      C(n) = -i\sqrt{n} (cosh(\theta(n))a(-n) - sinh(\theta(n))a^*(n))
$$
$$
       C^*(n) = i\sqrt{n} (cosh(\theta(n))a^*(-n) - sinh(\theta(n))a(n))
$$
$$
      D(n) = -i\sqrt{n} (cosh(\theta(n))a(n) - sinh(\theta(n))a^*(-n))
$$
$$
       D^*(n) = i\sqrt{n} (cosh(\theta(n))a^*(n) - sinh(\theta(n))a(-n))
$$
Here we use the c-number functions:
$$
 P_1 = \fr{n\pi},{L}\;\;\;\; P_0(n) = \sqrt{m^2 + P^2_1 (n)} \;\;\;\;\theta (n) = \ha \ln \fr{P_0 (n)},{|P_1(n)|} 
$$
For $n = 0$ we have:
$$
a(0) \equiv i\left(\sqrt{\fr{eL},{\sqrt{\pi}}}A_1 +
i\sqrt{\fr{\sqrt{\pi}L},{e}}\partial_0A_1 -
\fr{\pi^{\fr{3},{4}}},{2e^{\fr{3},{2}}\sqrt{L}}Q_5\right)
$$
$$
a^*(0) \equiv -i\left(\sqrt{\fr{eL},{\sqrt{\pi}}}A_1 -
i\sqrt{\fr{\sqrt{\pi}L},{e}}\partial_0A_1 -
\fr{\pi^{\fr{3},{4}}},{2e^{\fr{3},{2}}\sqrt{L}}Q_5\right)
$$
In the new variables the Hamiltonian is:
$$
H =  \fr{\pi},{4L e^2} Q^2 + \sum^{\infty}_{n=-\infty} P^0(n) a^*(n) a(n)
$$
{}From which we see that $\ket{\Omega}$ will be a ground state of the
system if:
$$
      a(n) \ket{\Omega} = 0
$$
and:
$$
      Q \ket{\Omega} = 0
$$
This last requirement is necessary for all states in the physical
subspace.

The similarity transformation which diagonalizes the Hamiltonian is:
$$
        SD(n)S^{-1} = i \sqrt{n} a(n) \qquad \; \qquad n > 0
$$
$$
        SC(-n)S^{-1} = i \sqrt{n} a(n) \qquad \; \qquad n < 0
$$
$$
       S\alpha_1S^{-1} = a(0)
$$
$$
            S = S_0S^\prime
$$

$$
S_0 = \exp\Bigl[\fr{\theta(0)},{2} (\alpha^2_1 - \alpha^{* 2}_1) -
\fr{\theta(0)},{e^{\theta(0)} -1 } \fr{Q_5},{2m \sqrt{Lm}} 
(\alpha_1 - \alpha^*_1)\Bigr]
$$

$$
S^\prime =
\exp\Bigl[ \sum_{n = 1}^\infty \fr{\theta(n)},{n} \left(C(n)D(n) -
D^*(n) C^*(n) \right)\Bigr]
$$
Since:
$$
          a(n) S\ket{M,N} = S(C or D) S^{-1} S\ket{M,N} = 0
$$
and:
$$
        SQS^{-1} = Q
$$
We see that any state of the form:
$$
              \ket{\Omega(M)} \equiv S\ket{M,-M}
$$
will be a ground state of the system. It is easy to show that:
$$
                   \bra{\Omega(M)}\bar{\psi}\psi\ket{\Omega(M)} = 0
$$
So if $\ket{\Omega(M)}$ is chosen for the ground state there is no
chiral condensate.  There is a subtle reason why that is not a good
choice however$^6$.  If we wish to have a solution which satisfies the
cluster property we must have the vacuum be an eigenstate of the mass
operator:
$$
    ( \psi_1^* \psi_2 + \psi_2^* \psi_1 ) \ket{\Omega} \sim
    \ket{\Omega}
$$ 
To do that we must choose the vacuum to be a $\theta$-state:
$$
\ket{\Omega(\theta)}  \equiv  \sum_{n=-\infty}^\infty 
e^{iM\theta}\ket{\Omega(M)}
$$

We can now calculate the chiral condensate.  We rewrite the fields in
terms of the new variables:
$$
   \psi_1^* \psi_2 = {1\over {2L}}{\ub\sigma}_1^*{\ub\sigma}_2 e^{
   \lambda_1^{(-)}} e^{ \lambda_1^{(+)}} e^{ - \lambda_2^{(-)}} e^{ -
   \lambda_2^{(+)}}
$$
Where:
$$
       \lambda_1^{(+)} = \sum_{n=1}^\infty {i\over
       \sqrt{n}}(cosh(\theta(n))a(n)e^{-ipx} -
       sinh(\theta(n))a^*(-n))e^{ipx}
$$
$$
    \lambda_1^{(-)} = \sum_{n=1}^\infty {i\over
    \sqrt{n}}(cosh(\theta(n))a^*(n)e^{ipx} -
    sinh(\theta(n))a(-n))e^{-ipx}
$$
$$
      - \lambda_2^{(+)} = \sum_{n=1}^\infty {{-i}\over \sqrt{n}}
      (cosh(\theta(n))a(-n)e^{-ipx} - sinh(\theta(n))a^*(n)e^{ipx})
$$
$$
    - \lambda_2^{(-)} = \sum_{n=1}^\infty {{-i}\over
    \sqrt{n}}(cosh(\theta(n))a^*(-n)e^{ipx} -
    sinh(\theta(n))a(n))e^{-ipx}
$$
If we now commute all the destruction operators forward and all the
creation operators backward, and use the relations:
$$
     {\ub\sigma}_1^*{\ub\sigma}_2 \ket{\Omega(M)} = \ket{\Omega(M+1)}
$$
$$
        {\ub\sigma}_1^*{\ub\sigma}_2 \ket{\Omega(\theta)} =
        e^{-i\theta}\ket{\Omega(\theta)}
$$
we find that:
$$
       \psi_1^* \psi_2 \ket{\Omega(\theta)} = {1\over {2L}}e^X
       e^{-i\theta}\ket{\Omega(\theta)}
$$
Where:
$$
e^X = \exp\Bigl[-2\sum_{n=1}^\infty{1\over n}(\sinh^2(\theta(n)) -
	\sinh(\theta(n))\cosh(\theta(n))) \Bigr]
$$
It is easy to show that:
$$
       \lim_{L\to\infty} e^X = \fr{m},{2\pi} e^\gamma L
$$
So we find that:
$$
             \lim_{L\to\infty}\bra{\Omega} \bar{\psi} \psi
             \ket{\Omega} = \fr{m},{2\pi} e^\gamma cos\theta
$$
We thus see that the fact that the chiral condensate survives to large
$L$ is due to the infrared divergence in the factor $e^X$; that factor
arises from the fact that there is mixing between the positive and
negative frequencies due to the interaction---it has the same source
as the wavefunction renormalization constant.

It will be important for future considerations to note that in this
solution Fermi products are regularized as:
$$
\normord{\psi^\dagger\psi}=\lim_{\e\ra0\atop \e^2<0}\Biggl\{
	e^{-ie\int_x^{x+\e} A_\nu^{(-)}dx^\nu}
	\psi^\dagger(x+\e)\psi(x)e^{-ie\int_x^{x+\e}
	A_\nu^{(+)}dx^\nu} -{\rm V.E.V.}\Biggr\}
$$
That is, with a gauge invariant splitting which can be on the
spacelike initial value surface $t = 0$; as in the case of free
theory, however, a splitting in the lightlike direction $x^-$ is
allowed.  We shall return to this point later.

\section{ On the Light-Cone }

Let us now contrast the above results with those obtained$^3$ when the
same Lagrangian is quantized on characteristic surfaces: $\psi_+$ on
$x^+ = 0$;$\psi_-$ on $x^- =0$, with antiperiodic boundary conditions
in each case.  We use the gauge $\partial_-A^+=0$.  We initialize the
fields to be isomorphic to free fields on these surfaces:
$$
\psi_+ = {1\over \sqrt{2L}}
e^{\lambda_+^{(-)}(x)} \sigma_+(x) e^{\lambda_+^{(+)}(x)}
$$
$$
\psi_- = {1\over\sqrt{2L}} e^{-\lambda_D^*(x^+)}
\sigma_-(x)e^{\lambda_D(x^+)}        
$$
where:
$$
\lambda_+(x) = -i\sqrt{{\pi\over L}}\sum_{n=1}^\infty
{1\over \sqrt{p_-(n)}}\left( C(n)e^{-ip(n)x} + C^*(n)e^{ip(n)x}\right)
$$
$$
\lambda_D(x^+) = \sum_{n=1}^\infty{1\over\sqrt{n}} D(n)e^{-ik_+(n)x^+} 
$$
In this case there is a zero mode in each component of the gauge field
but neither is a degree of freedom.  These zero modes are found to be:
$$
   A^+=-{1\over Lm^2}Q_-  \quad ;\quad  A^-(0)=-{1\over Lm^2}Q_+    
$$
The operator which controls the dynamics, $P^-$, is found to be:
$$
P^- = {1\over 4Lm^2}(Q_-^2-Q_+^2)+\sum_{n=1}^\infty p^-(n) C^*(n) C(n)
  + \sum_{n=1}^\infty 2k_+(n) D^*(n)D(n)
$$
where $Q_+$ and $Q_-$ are respectively the charge in the $\psi_+$ or
$\psi_-$ field and we have used the c-number functions:
$$
    P^-(n) = {{m^2L}\over {2n\pi}}\quad ;\quad k_+(n) =
    {{n\pi}\over{L}}
$$
Thus the Hamiltonian and the momentum are already diagonal and there
is no need for an S-operator.

Arguments essentially similar to those we used in Coulomb gauge tell
us that if we define:
$$
             \ket{\Omega(M)} = \ket{M,-M}
$$
the ground state of the system will be:
$$
\ket{\Omega(\theta)}  \equiv  
\sum_{n=-\infty}^\infty e^{iM\theta}\ket{\Omega(M)}
$$
An easy calculation gives:
$$
\psi_+^* \psi_- \ket{\Omega(\theta)} = 
{1\over {2L}} e^{-i\theta}\ket{\Omega(\theta)}
$$
so the chiral condensate is:
$$
\bra{\Omega} \bar{\psi} \psi \ket{\Omega} = \fr{1},{L} cos\theta ,
$$
which in the large $L$ limit goes to 0. The reason the chiral
condensate vanishes in the large $L$ limit, in contrast to the previous
section, is the lack of the factor $e^X$.  That factor arose due to
the mixing of the positive and negative frequency modes and its
absence here is due to the lack of such mixing.  The fact that the
positive and negative frequency modes do not mix in this gauge with
these boundary conditions was first discussed by Eller, Pauli and
Brodsky$^7$.  For the exact same reason there is no wave function
renormalization.  Note here that we may define Fermi products as:
$$
\normord{\psi_+^*(x) \psi_+(x)} \equiv 
\quad \lim_{\epsilon^-\rightarrow0} \left(
e^{-ie\int_x^{x+\epsilon^-} A_-^{(-)} dx^-}
\psi_+^*(x+\epsilon^-) \psi_+(x)
e^{-ie\int_x^{x+\epsilon^-} A_-^{(+)}dx^-}-{\rm V.E.V.}\right)
$$
Splittings in spacelike directions are also allowed, but the point is
that, just as in free theory, splitting in the lightlike initial value
surface is allowed.
  
We might wonder whether the differences between the two solutions we
have discussed are due to the choice of gauge or the choice of
boundary conditions.  Let us briefly consider the solution$^5$ in
light-cone gauge initialized on the surface $t = 0$ with antiperiodic
boundary conditions on the Fermi fields and periodic boundary
conditions on the gauge fields ( the field $A^-$ is no longer a
constraint ).  The Fermi fields are initialized as in the case of the
Coulomb gauge and again diagonalization of the Hamiltonian requires
mixing of the negative and positive frequency modes.  The
diagonalization is accomplished in terms of a massive pseudoscalar
field, $\tilde{\Sigma}$, half the degrees of freedom of a massless
ghost, $\tilde{\eta}$ and half the degrees of freedom of a massless
scalar, $\lambda_1$:
$$
          \psi_- = \psi_-(FREE) =
          {1\over\sqrt{2L}}e^{-\lambda_1^{(-)}(x)}\sigma_1(x)
          e^{-\lambda_1^{(+)}(x)}
$$
$$
     \psi_+ = e^{[-2i\sqrt{\pi}(\tilde{\eta}^{(-)} +
     \tilde{\Sigma}^{(-)})]} \sigma_+
     e^{[-2i\sqrt{\pi}(\tilde{\eta}^{(+)} + \tilde{\Sigma}^{(+)})]}
$$
The ghost field plays an essential role in regulating Fermi products.
We have:
$$
\langle  e^{[-2i\sqrt{\pi} \tilde{\Sigma}^{(+)}(x + \epsilon)]} 
e^{[-2i\sqrt{\pi} \tilde{\Sigma}^{(-)}(x)]} \rangle \approx e^X 
\fr{-i},{2\pi\e^-} \fr{-i},{2\pi\e^+}
$$
while:
$$
\langle  
e^{[-2i\sqrt{\pi} \tilde{\eta}^{(+)}(x + \epsilon)]} e^{[-2i\sqrt{\pi}
\tilde{\eta}^{(-)}(x)]} \rangle \approx (\fr{-i},{2\pi\e^+})^{-1}
$$
We thus find that the proper working of the gauge invariant point
splitting regulator requires that we split in a spacelike direction
and have the ghost field present.  The $x^+$ singularity in the first
case comes from the small $P^-$ region of momentum space, while in the
second case it comes from the large $P^-$ region.  The coulomb gauge
solution split the singularities in $x^-$ and $x^+$ between the
$\psi_+$ and $\psi_-$ fields just as in free theory, but in light-cone
gauge all the dynamics is placed on the $\psi_+$ field and both
singularities arise, one of them canceled by the ghost.  In light-cone
gauge with antiperiodic boundary conditions on $x^+ = 0$, the small
$P^-$ region of momentum space is removed and the $x^+$ singularity
does not arise.

The solution$^5$ just discussed goes smoothly into the light-cone
gauge continuum solution of Nakawaki$^4$. In the continuum case the
solution is also written in terms of a massive pseudoscalar, a
massless ghost and a massless scalar.  In fact one can almost get from
the periodic solution to the continuum one by changing the dispersion
relation for all fields in the periodic solution to the relevant
continuum one.  The extra differences are that a more complicated
spurion is needed in the continuum case and a Klaiber$^8$ regulator is
needed to control the infrared.  One still finds that the splitting
must be in a strictly spacelike direction with the ghost field
cancelling out an unwanted singularity just as in the periodic case.
That means that while Nakawaki's solution can be quantized on either
$t = 0$ or $x^+ = 0$, if the characteristic surface is used then
dynamics gets involved in the definition of the operators and there
arises extra subtleties.

\section{ Remarks }

All the solutions in light-cone gauge have fields which are functions
of $x^+$ and which are essential parts of the solution.  Thus, in that
gauge one must use either a spacelike surface or both characteristic
surfaces, $x^+$ and $x^-$, to properly initialize the problem.

The light-cone gauge solutions which have:
$$
\bra{\Omega} \bar{\psi} \psi \ket{\Omega} \neq 0
$$
in the limit as the regulator is removed have the property that point
splitting must be done on a spacelike surface.  This fact gives rise
to extra complications in quantizing these solutions on the
characteristic surfaces in that the definition of the operator
products must involve points outside the initial value surface and
thus dynamics gets involved in defining the dynamical generators.  It
is not clear how this complication carries over to higher dimensions.
The functions of $x^+$ are present in gauge theories in all dimensions
but a nonperturbative regulator for defining field products, with a
procedure for renormalizing these products is not known.  Thus, even
though the field products could be split in a spacelike direction
within the initial value surface for theories in dimensions higher
than two, the implications of this fact for a possible kinematical
definition of the field products is not clear.

If we impose periodicity conditions on the characteristic surfaces,
the operator products can be regulated by splitting in the initial
value surface.  This greatly simplifies the technical formulation of
the theory but has the property that some of the subtle details of the
operator products, such as the condensate, do not approach their
continuum values as the periodicity length is taken to infinity.
However, the light-cone gauge solution with periodicity conditions on
the characteristics gives the correct spectrum and the correct
s-matrix while having a much simpler ( though not completely trivial )
vacuum and a much simpler solution.  The anomaly is also correctly
given.  The degree to which one may lose the ability to represent some
aspects of the physics in order to obtain the simplifications derived
from a kinematic definition of the field products is not entirely
understood.  It would be very valuable to be able to make more general
statements on this subject.

The procedure of introducing periodicity conditions on the
characteristic surfaces is very similar to the regulator used by 't
Hooft$^9$ to solve large-N QCD in two dimensions.  Both have the
effect of simply removing the small $P^+$ region without inducing any
counterterms.  That solution is known to have an inconsistency
involving the condensate: The 't Hooft$^9$ propagator has zero
condensate but the spectrum derived from it implies a nonzero value
for the condensate$^{10}$. I believe that the features of the
Schwinger model solutions we have been discussing will be found
relevant to resolving that apparent inconsistency in the solution to
the 't Hooft model.

\end{document}